\documentstyle[12pt]{article}
\newcounter{eqn}
\def\lab{\refstepcounter{eqn}\eqno(\arabic{eqn})}
\def\l#1{\lab\label{#1}}
\def\r#1{(\ref{#1})}
\def\c#1{\cite{#1}}

\begin{document}

\begin{titlepage}
\begin{flushright}
July 9, 2001\\
hep-th/0107064
\end{flushright}

\begin{centering}
\vfill
{\bf 
REFINED ALGEBRAIC  QUANTIZATION  OF CONSTRAINED SYSTEMS WITH STRUCTURE
FUNCTIONS
}
\vspace{1cm}

O. Yu. Shvedov \footnote{shvedov@qs.phys.msu.su} \\
\vspace{0.3cm}
{\small {\em Sub-Dept. of Quantum Statistics and Field Theory,}}\\
{\small{\em Department of Physics, Moscow State University }}\\
{\small{\em Vorobievy gory, Moscow 119899, Russia}}

\vspace{0.7cm}

\end{centering}

{\bf Abstract}

The method of refined algebraic quantization  of  constrained  systems
which is  based  on  modification  of  the inner product of the theory
rather than  on  imposing  constraints  on  the  physical  states   is
generalized to   the   case  of  constrained  systems  with  structure
functions and open  gauge  algebras.  A  new  prescription  for  inner
product for  the open-algebra systems is suggested.  It is illustrated
on a simple example.  The  correspondence  between  refined  algebraic
and BRST-BFV  quantizations is investigated for the case of nontrivial
structure functions.

\vspace{0.7cm}

\vfill \vfill
\noindent

\end{titlepage}
\newpage

\section{Introduction}

Theory  of constrained systems is a basis of modern physics:
gauge field theories,  quantum gravity and  supergravity,  string  and
superstring models are examples of systems with constraints.  For such
theories, one should  specify  not  only  an  evolution  equation  but
additional requirements  (constraints) imposing on initial conditions.
For some cases,  Hamiltonian is zero,  so that  all  the  dynamics  is
involved to  constraints,  and  the well-known problem of time in
reparametrization-invariant theories arises.

In quantum mechanics of constraint systems evolution  (if  exists)  is
taken into  account  in  a  standard way:  evolution transformation is
presented as  $\exp(-iH^+t)$  for  some  Hamiltonian   $H$.   However,
constraints can be taken into account in different ways.

First, one  can  use  the  original  Dirac idea \c{D} and consider the
states to obey the following additional conditions:
$$
\hat{\Lambda}_a^+ \Psi = 0, \qquad a=\overline{1,M},
\l{1}
$$
where $\hat{\Lambda}_a^+$   are   quantum   analogs   of  constraints.
Requirements \r{1} do not contradict each  other,  provided  that  the
constraints commute  on the constraint surface.  For the quantum case,
this means that
$$
[\hat{\Lambda}_a^+, \hat{\Lambda}_b^+]      =      i      (U^c_{ab})^+
\hat{\Lambda}_c^+
\l{2}
$$
for some operators $U^c_{ab}$ called usually  as  structure  functions
\c{Henneaux}. Relation   \r{1}   should   also   conserve  under  time
evolution, so that constraints should commute with the Hamiltonian for
the states obeying eq.\r{1}:
$$
[H^+, \hat{\Lambda}_a^+] = i (R^c_a)^+ \hat{\Lambda}_c^+
\l{3}
$$
for some operators $R^c_a$.

The most  difficult  problem  of the Dirac approach is construction of
the inner product,  since  $\Psi(q)$  are  distributions  rather  than
square-integrable functions.  One  usually  imposes  additional  gauge
conditions \c{D,Henneaux,Faddeev} in such a way that each gauge  orbit
should be  taken  into account once.  Unfortunately,  this approach is
gauge-dependent, especially for the case of the Gribov copies  problem
\c{Gribov,Shabanov}.

Therefore, other  ways  of  quantization  of  constrained systems were
invented. The BRST-BFV approach \c{BFV} is based on extension  of  the
phase space   by  introducing  Lagrange  multipliers  $\lambda^a$  and
momenta $\pi_a$,  ghosts  and  antighosts   $C^a$,   $\overline{C}_a$,
momenta $\overline{\Pi}_a$,  $\Pi^a$.  A manifestly covariant operator
formulation of nonabelian gauge theories was obtained in this approach
\c{KO}.

An alternative way to develop the quantum theory is to use the refined
algebraic quantization  technique  \c{Marolf}  and  modify  the  inner
product instead  of imposing requirements \r{1}.  States are specified
by arbitrary functions $\Phi(q)$ called auxiliary state  vectors,  but
their inner product is given by the nontrivial formula
$$
(\Phi, \eta \Phi).
\l{4}
$$
For the abelian case, $\eta \sim \prod_a \delta(\hat{\Lambda}_a)$. For
the general case, the main requirement for the operator $\eta$ ($\eta
= \eta^+$) is the following
$$
\eta \hat{\Lambda}_a = 0.
\l{5}
$$
The positivity  condition  is  also  imposed.  Two  states  are called
equivalent if their difference $\Delta \Phi$ has zero  norm.  This  is
certainly the case if
$$
\Delta \Phi = \hat{\Lambda}_a Y^a
$$
because of  \r{5}.  The classes of equivalence are identified with the
Dirac states with the help of formula
$$
\Psi = \eta \Phi.
\l{6}
$$
The constrained conditions \r{1} are automatically satisfied then. The
inner product for physical states is introduced.

Analogous ideas   were   used  in  the  projection  operator  approach
\c{proj,Shabanov}.

Quantum observables  $H$  can  be  introduced  in  this  approach   as
operators in  the  space  of  auxiliary  states $\Phi$.  The
unitarity property is written as follows
$$
(e^{-iHt})^+ \eta e^{-iHt} = \eta.
$$
This means that
$$
H^+\eta = \eta H.
\l{6a}
$$
Let $\Phi$  be  an  auxiliary  state  corresponding to the Dirac state
$\Psi = \eta \Phi$.  The observable $H$  takes  it  to  $H\Phi$.  This
corresponds to the Dirac state $\eta H\Phi = H^+ \eta \Phi = H^+\Psi$.
Therefore, it is the operator $H^+$ that corresponds to the observable
$H$ in  the  Dirac  approach,  while  $\exp(-iH^+t)$  is  an evolution
operator. Thus,  an observable in the refined  algebraic  quantization
approach is an operator satisfying properties \r{3}, \r{6a}.

An explicit form of the operator $\eta$ has been obtained only for the
case of closed algebra of constraints,  where $U^c_{ab} = const$.  For
this case,  $\eta$  is  expressed via the integral over gauge group of
the representation  of  the  constraint  group  \c{M2}.  It  has  been
stressed in \c{Marolf3} that generalization of this formula to systems
with structure functions $U^c_{ab} \ne const$ is an  interesting  open
problem. The  purpose  of this paper is to write down the corresponding
prescription for the inner product.

\section{A proposal for the inner product bilinear form}

It is convenient to use another  well-developed  quantization
technique, the    BRST-BFV   approach   \c{BFV,Henneaux}.   Additional
operators $\lambda^a$,  $\pi_a$,  $C^a$,  $\overline{C}_a$,   $\Pi^a$,
$\overline{\Pi}_a$ obeying   the   following   nontrivial  commutation
relations
$$
[\lambda^a, \pi_b] = i\delta^a_b,  \qquad  [C^a,  \overline{\Pi}_b]_+  =
\delta^a_b, \qquad [\overline{C}_a, \Pi^b]_+ = \delta_a^b
$$
are introduced.    Operators    $\overline{C}_a$   and   $\Pi^b$   are
anti-Hermitian. The main object is the B-charge which is  a  nilpotent
Hermitian operator looked for in the following form
$$
\Omega =   -   i\pi_a   \Pi^a   +   C^a   \hat{\Lambda}_a   +   ...  +
\Omega^n{}^{b_1...b_{n-1}}_{a_1...a_n} \overline{\Pi}_{b_1}        ...
\overline{\Pi}_{b_{n-1}} C^{a_1} ... C^{a_n} + ...
\l{7}
$$
The operators $\overline{\Pi}$ and $C$ are ordered in formula \r{7} in
such a way that ghosts $C$ are put to the right, while the momenta
$\overline{\Pi}$ are put to the left.  The property  $\Omega^+=\Omega$
means that in general
$$
\hat{\Lambda}^+ \ne \hat{\Lambda}.
\l{8}
$$
The operator-valued coefficient functions
$\Omega^n{}^{b_1...b_{n-1}}_{a_1...a_n}$
being antisymmetric separately with respect to  $b_1,...,b_{n-1}$  and
separately with respect to $a_1,...,a_n$ are constructed in a standard
way \c{Henneaux} from recursive relations that are corollaries of  the
property $\Omega^2=0$.

Physical states in the BRST-BFV approach satisfy the requirement
$$
\Omega\Upsilon = 0,
\l{9}
$$
while the transformation
$$
\Upsilon \to \Upsilon + \Omega Y
\l{10}
$$
is gauge: states $\Omega Y$ are equivalent to zero.

To specify an inner product  for  BRST-BFV  quantization  is  also  a
difficult problem (see,  for example, \c{Razumov,Marnelius}). The most
general approach was suggested in \c{Marnelius}.  One considers  such
physical states that
$$
\pi_a \Upsilon  =0,  \qquad  C^a  \Upsilon = 0,  \qquad \overline{C}_a
\Upsilon =0.
\l{11}
$$
Such states   are   automatically   B-invariant:  condition  \r{9}  is
satisfied. Since the inner product $(\Upsilon,\Upsilon)$ contains  the
factor like $\infty \cdot 0$, one uses another prescription
$$
(\Upsilon, e^{t[\Omega,\rho]_+} \Upsilon)
\l{12}
$$
being formally  equivalent  to   $(\Upsilon,\Upsilon)$   but   without
divergences. It is convenient to choose the gauge fermion $\rho$ as
$$
\rho = - \lambda^a \overline{\Pi}_a
$$
so that
$$
[\Omega,\rho]_+ = - \overline{\Pi}_a \Pi^a - \lambda^a \hat{\Omega}_a
$$
with
$$
\hat{\Omega}_a =   \Omega_a  (\overline{\Pi},C)  =  [\overline{\Pi}_a,
\Omega]_+ = \hat{\Lambda}_a + ... + n
\Omega^n{}^{b_1...b_{n-1}}_{a_1...a_{n-1}a}
\overline{\Pi}_{b_1}
...
\overline{\Pi}_{b_{n-1}}
C^{a_1}... C^{a_{n-1}} + ...
$$

It is   convenient   to   introduce    the    functional    Schrodinger
representation. $B$-states   $\Upsilon$  are  specified  as  functions
$\Upsilon(q,\lambda,\Pi,\overline{\Pi})$. The  inner   product   reads
\c{JMP0}
$$
(\Upsilon_1,\Upsilon_2) =   \int   dq
\prod_{a=1}^M   d\mu_a   d\overline{\Pi}_a d\Pi^a
(\Upsilon_1(q,i\mu,\Pi,\overline{\Pi}))^*
\Upsilon_2(q,-i\mu,\Pi,\overline{\Pi})
\l{14}
$$
where $(\Pi^a)^* = - \Pi^a$,  $\overline{\Pi}_a^* = \overline{\Pi}_a$.
The ghost  momenta  and   bosonic   coordinates   are   presented   as
multiplicators $\Pi^a$ and $\overline{\Pi}_a$, while
$$
C^a = \frac{\partial}{\partial \overline{\Pi}_a}, \qquad
\overline{C}_a = \frac{\partial}{\partial \Pi^a}, \qquad
\pi_a = - i \frac{\partial}{\partial \lambda^a}, \qquad
p_i = - i \frac{\partial}{\partial q^i}
\l{14a}
$$
(where the left derivatives are considered). Conditions \r{11} mean
$$
\Upsilon= \Phi(q).
$$
It has been argued in \c{Shv} that $\Phi(q)$ should be identified with
the auxiliary state in the refined  algebraic  quantization  approach.
Formulas \r{12} and \r{14} give us the following inner product,
$$
\int dq \Phi^*(q) \prod_{a=1}^M d\mu_a d\overline{\Pi}_a d\Pi^a
e^{-t\overline{\Pi}_a \Pi^a + it\mu_a \hat{\Omega}_a} \Phi(q)
\l{15}
$$
with $\hat{\Omega}_a  =
\Omega_a  (\overline{\Pi},  \partial/\partial \overline{\Pi})$.
Formula \r{15} is of the type \r{4} with
$$
\eta = \int \prod_{a=1}^M d\mu_a d\overline{\Pi}_a d\Pi^a
e^{-\overline{\Pi}_a \Pi^a + i\mu_a \hat{\Omega}_a
(\overline{\Pi},  \partial/\partial \overline{\Pi})} 1.
\l{16}
$$
Here $\Pi^a$ and $\mu_a$ are rescaled in $t$ times.  Formula \r{16} is
a generalization  of  Giulini-Marolf formula of \c{M2} for the case of
open algebras,  since for the Lie-algebra case results of  \c{M2}  are
reproduced by formula \r{16} \c{Shv}.

One should  take  into account the topological problems analogously to
the closed-algebra case \c{Shv}: integration may be performed not over
all values of $\mu$ but over $\mu$ belonging to some domain.

\section{Properties of the inner product}

Let us investigate properties of the operator $\eta$ \r{16}.
First of all,  check that $\eta^+ = \eta$, so that formula \r{4} gives
us real values. One has
$$
(\Phi, \eta  \Phi)  =  \int  \prod_{a=1}^M  d\mu_a  (\Phi,  \exp[\Pi^a
\overline{\Pi}_a + i \mu_a \hat{\Omega}_a] \Phi)
$$
and
$$
(\Phi, \eta \Phi)^* = \int  \prod_{a=1}^M  d\mu_a  (\Phi,
\exp[\Pi^a \overline{\Pi}_a - i \mu_a \hat{\Omega}_a^+] \Phi)
$$
After change  of  variables $\mu_a \to - \mu_a$ and using the property
$\hat{\Omega}_a^+ = \hat{\Omega}_a$ being a corollary of the relations
$\overline{\Pi}_a^+ =  \overline{\Pi}_a$  and $\Omega^+ = \Omega$,  we
find $\eta^+ = \eta$.

Let us check relation \r{5}. One has
$$
\eta \hat{\Lambda}_b Y^b(q) = \int
\prod_{a=1}^M d\mu_a d\overline{\Pi}_a d\Pi^a
\exp[\Pi^a \overline{\Pi}_a   +   i   \mu_a   \hat{\Omega}_a]   \Omega
\overline{\Pi}_b Y^b(q).
\l{17}
$$
Since $\Omega^2=\Omega$,  the operators $\Omega$ and  $[\Omega,\rho]_+$
commute:
$$
\Omega [\Omega,\rho]_+ = \Omega \rho \Omega = [\Omega,\rho]_+ \Omega,
\l{17a}
$$
so that
$$
e^{\Pi^a \overline{\Pi}_a + i\mu_a \hat{\Omega}_a} \Omega =
\Omega e^{\Pi^a \overline{\Pi}_a + i\mu_a \hat{\Omega}_a}.
$$
Formula \r{17} transforms then to
$$
\eta \hat{\Lambda}_b Y^b(q) = \int
\prod_{a=1}^M d\mu_a d\overline{\Pi}_a d\Pi^a
\Omega^+
\exp[\Pi^a \overline{\Pi}_a   +   i   \mu_a   \hat{\Omega}_a]
\overline{\Pi}_b Y^b(q)
\l{18}
$$
since $\Omega = \Omega^+$. The operator $\Omega^+$ can be presented in
representation \r{14a} as
$$
\Omega^+ =      \frac{1}{i} \frac{\partial}{\partial \mu_a}   \Pi^a   +
\frac{\partial}{\partial \overline{\Pi}_a}   \hat{\Lambda}_a^+   +
...  +
(\Omega^n{}^{b_1...b_{n-1}}_{a_1...a_n})^+
\frac{\partial}{\partial \overline{\Pi}_{a_n}} ...
\frac{\partial}{\partial \overline{\Pi}_{a_1}}
\overline{\Pi}_{b_{n-1}}
... \overline{\Pi}_{b_1} + ...
\l{BRST}
$$
Integral \r{18} then vanishes  as  an  integral  of  full  derivative.
Formula \r{5} is checked.
Thus, formula \r{16} obeys the  desired  properties  of  the  operator
$\eta$ entering   to  the  inner  product.  However,
the problem of positive definiteness of the inner product  remains  to
be investigated.

\section{Correspondence between BRST-BFV,  Dirac and refined algebraic
quantization approaches}

Let us show that  correspondence  between  BFV,  auxiliary  and  Dirac
states found in \c{Shv} for the Lie-algebra case remains valid for the
case of nontrivial structure functions.

Let $\Upsilon$ be an arbitrary BFV  state  obeying  eq.\r{9}  but  not
satisfying in general eq.\r{11}. For this BFV state,
consider the function
$$
\Phi(q) = \Upsilon(q,0,0,0)
\l{refined}
$$
It occurs  to play a role of an auxiliary wave function in the refined
algebraic quantization approach.  If the conditions \r{11} are  valid,
relation \r{refined}  is  obvious.  It  was advocated in \c{Marnelius}
that any  physical  state  can  be  taken  to  the  gauge  \r{11}   by
transformation \r{10}. However, the auxiliary state
$$
(\Omega Y)(q,0,0,0)    =    \hat{\Lambda}_a   \frac{\partial}{\partial
\overline{\Pi}_a}|_{\overline{\Pi}=\Pi=0, \lambda = 0} Y
$$
is equivalent to zero (has zero norm) because of \r{5}. Therefore, for
any $\Upsilon$   definition  \r{refined}  is  valid  since  equivalent
auxiliary states are identified.

Eq.\r{6} for the Dirac wave function can be rewritten as
$$
\Psi(q) =   \int   \prod_{a=1}^M   d\mu_a   d\overline{\Pi}_a   d\Pi^a
(e^{[\Omega,\rho]_+} \Phi)(q, -i\mu, \Pi, \overline{\Pi}).
$$
Let $\Upsilon$  be an arbitrary BFV state that is equivalent to $\Phi$
(and, therefore, to $e^{[\Omega;\rho]_+}\Phi$). Let us check that
$$
\Psi(q) = \int   \prod_{a=1}^M   d\mu_a   d\overline{\Pi}_a   d\Pi^a
\Upsilon (q, -i\mu, \Pi, \overline{\Pi}).
\l{Dirac1}
$$
It is sufficient to justify that equivalent  BFV  states  \r{10}  give
equal Dirac  wave functions \r{Dirac1}.  However,  it follows directly
from \r{BRST} that
$$
\int   \prod_{a=1}^M   d\mu_a   d\overline{\Pi}_a   d\Pi^a
(\Omega^+ Y) (q, -i\mu, \Pi, \overline{\Pi}) = 0
$$
since integrals  of  full  derivatives  vanish.  Formula \r{Dirac1} is
obtained.

To check relation \r{1}, use the property
$$
\int   \prod_{a=1}^M   d\mu_a   d\overline{\Pi}_a   d\Pi^a
(\Omega^+ \overline{\Pi}_a \Upsilon)
(q, -i\mu, \Pi, \overline{\Pi}) = 0.
$$
Since $\Omega^+\Upsilon  =  \Omega\Upsilon =0$,  one can rewrite it as
follows,
$$
\int   \prod_{a=1}^M   d\mu_a   d\overline{\Pi}_a   d\Pi^a
([\Omega^+, \overline{\Pi}_a]_+ \Upsilon)
(q, -i\mu, \Pi, \overline{\Pi}) = 0.
\l{i1}
$$
The anticommutator has the form
$$
[\Omega^+;\overline{\Pi}_a]  =
\hat{\Lambda}_a^+   +
...  +
n (\Omega^n{}^{b_1...b_{n-1}}_{a_1...a_{n-1}a})^+
\frac{\partial}{\partial \overline{\Pi}_{a_{n-1}}} ...
\frac{\partial}{\partial \overline{\Pi}_{a_1}}
\overline{\Pi}_{b_{n-1}}
... \overline{\Pi}_{b_1} + ...
$$
It contains full derivatives, except for the term $\hat{\Lambda}_a^+$.
Thus, eq.\r{i1} can be presented as
$$
\hat{\Lambda}_a^+
\int   \prod_{a=1}^M   d\mu_a   d\overline{\Pi}_a   d\Pi^a
\Upsilon (q, -i\mu, \Pi, \overline{\Pi}) = 0.
$$
Eq.\r{1} is obtained.

Thus, the  {\it  formal  }  relationship  between  refined   algebraic
quantization, Dirac  and  BFV states is found.  However,  the topology
problems which may lead to integration over some domain in  \r{Dirac1}
are to be investigated in future.

\section{Quantum observables}

Let us consider the properties of quantum observables. In the
BRST-BFV approach, observables are viewed as series
$$
H_B = H + ... +
H^n{}^{b_1...b_{n}}_{a_1...a_n} \overline{\Pi}_{b_1}               ...
\overline{\Pi}_{b_n} C^{a_1} ... C^{a_n} + ...
$$
The operator  coefficient  functions  $H^n{}^{b_1...b_{n}}_{a_1...a_n}
(\hat{p},\hat{q})$
are chosen in such a way that
$$
H_B^+ = H_B, \qquad [\Omega, H_B] = 0.
\l{ham}
$$
These properties  provide  that physical states \r{9} are taken by the
operator $H$  to  physical,  while  equivalent  states  are  taken  to
equivalent.

Since $(H_B\Upsilon)(q,0,0,0) = H\Upsilon(q,0,0,0)$,
the  operator coefficient $H$ is an observable in the
refined algebraic quantization approach. One also has
$$
\int \prod_{a=1}^M     d\mu_a     d\overline{\Pi}_a    d\Pi^a    H_B^+
\Upsilon(q,-i\mu,\Pi,\overline{\Pi})
= H^+ \int \prod_{a=1}^M     d\mu_a     d\overline{\Pi}_a    d\Pi^a
\Upsilon(q,-i\mu,\Pi,\overline{\Pi})
$$
because integral of full derivative vanishes.  Therefore,  $H^+$ is  a
Hamiltonian in the Dirac approach.

To check property \r{3}, consider the expressions
$$
\Omega H_B   \overline{\Pi}_c   \Phi(q)|_{\overline{\Pi}=\Pi  =  0}  =
(\hat{\Lambda}_c H + \hat{\Lambda}_b H^1{}_c^b) \Phi(q)
$$
and
$$
H_B \Omega  \overline{\Pi}_c  \Phi(q)|_{\overline{\Pi}=\Pi  =   0}   =
H\hat{\Lambda}_c \Phi(q).
$$
They should be equal because of \r{ham},  so that $[H;\hat{\Lambda}_c]
= \hat{\Lambda}_b  H^1{}_c^b$  and eq.\r{3} is satisfied if $-iR^c_a =
H^1{}^c_a$.

Let us verify formula \r{6a}. One has
$$
\eta H \Phi(q) = \int
\prod_{a=1}^M d\mu_a d\overline{\Pi}_a d\Pi^a
\exp[\Pi^a \overline{\Pi}_a   +   i   \mu_a   \hat{\Omega}_a]
H_B \Phi(q),
\l{19}
$$
while
$$
H^+ \eta  \Phi(q) = \int
\prod_{a=1}^M d\mu_a d\overline{\Pi}_a d\Pi^a
H_B^+
\exp[\Pi^a \overline{\Pi}_a   +   i   \mu_a   \hat{\Omega}_a]
\Phi(q).
\l{20}
$$
Here we  have  taken  into account that $C^a \Phi(q) = 0$ and that the
integral of full  derivative  vanishes.  Consider  the  difference  of
eqs.\r{19}, \r{20}. Let us make use of the following relation,
$$
H_B^+ e^{[\Omega,\rho]_+} -  e^{[\Omega,\rho]_+} H_B = \int_0^1 d\tau
e^{\tau[\Omega,\rho]_+} [[\Omega,\rho]_+,       H_B]       e^{(1-\tau)
[\Omega,\rho]_+},
$$
since $H_B^+    =    H_B$.    Moreover,    $[[\Omega,\rho]_+,H_B]    =
[\Omega,[H_B,\rho]]_+$. It follows from eq.\r{17a} that
$$
H_B^+ e^{[\Omega;\rho]_+} - e^{[\Omega;\rho]_+} H_B = [\Omega;A]_+
$$
with
$$
A= \int_0^1 d\tau
e^{\tau[\Omega,\rho]_+} [H_B; \rho]       e^{(1-\tau)
[\Omega,\rho]_+},
$$
Therefore, the difference between formulas \r{19} and \r{20} reads
$$
(H^+\eta - \eta H)\Phi(q) = \int
\prod_{a=1}^M d\mu_a d\overline{\Pi}_a d\Pi^a
[\Omega^+A + A\Omega] \Phi(q).
$$
This integral vanishes since $\Omega \Phi(q) = 0$ and  an  integral  of
full derivative is zero. Thus, relation \r{6a} is satisfied.

\section{A simple example}

Consider a simple  example  of  a  system  with  structure  functions.
Investigate the   model   with   3  degrees  of  freedom  $(p_i,q^i)$,
$i=\overline{1,3}$ and 2 classical constraints
$$
\Lambda_1 = a(q^2,q^3) p_1, \qquad \Lambda_2 = p_2.
\l{20a}
$$
Since $\{\Lambda_1,   \Lambda_2\}   =   \partial_2   \log   a(q^2,q^3)
\Lambda_1$, the constraints forms an algebra with structure functions.
Let us  look for the B-charge in the form \r{7}.  In classical theory,
it should be written as
$$
\Omega = -i\pi_1 \Pi^1 - i \pi_2 \Pi^2  +  p_1  a  C^1  +  p_2  C^2  +
(\alpha_1 \overline{\Pi}_1 + \alpha_2 \overline{\Pi}_2) C^1C^2
\l{21}
$$
for some functions $\alpha_a(p,q)$.  The property $\{\Omega,\Omega\} =
0$ means that
$$
p_1 a \alpha_1 + p_2\alpha_2 = [p_1a; p_2],
$$
so that
$$
\alpha_1 = - i \partial_2 \log a; \qquad \alpha_2 = 0.
$$
We see that classically
$$
\Omega = -i\pi_1 \Pi^1 - i\pi_2\Pi^2 + p_1aC^1 + p_2C^2 - i \partial_2
\log a \overline{\Pi}_1 C^1C^2.
$$
To quantize the B-charge,  one should choose the operator ordering. If
the $\overline{\Pi}$-operators  were  put  to the left with respect to
$C$-operators, the quantum B-charge would be not  Hermitian.  To  obey
the condition $\Omega^+=\Omega$, let us use the Weyl quantization
$$
\Omega =   -i\pi_1\Pi^1   -   i   \pi_2\Pi^2   +   p_1aC^1  +  (p_2  -
i\overline{\Pi}_1 \partial_2 \log a C^1 + \frac{i}{2} \partial_2  \log
a)C^2
\l{22}
$$
It is   remarkable   that   in   quantum   theory    the    constraint
$\hat{\Lambda}_2$ should  be  modified  with  respect to the classical
theory \r{20a}; it follows from eq.\r{7} that
$$
\hat{\Lambda}_2 = p_2 + \frac{i}{2} \partial_2 \log a,
$$
so that the operator $\hat{\Lambda}_2$ becomes formally non-Hermitian.
This feature  of  quantum  constraints  is  known  from  the theory of
constrained systems with nonunimodular closed algebra \c{KS,M2}.

Let us evaluate the inner product \r{15}. Consider the wave function
$$
\Upsilon^t(q, \overline{\Pi},  \Pi) = e^{- t\overline{\Pi}_a \Pi^a +  it
\mu_a \hat{\Omega}_a} \Phi(q).
\l{23}
$$
Since
$$
\hat{\Omega}_1 = p_1a + i\overline{\Pi}_1 \partial_2 \log a C^2,
\qquad \hat{\Omega}_2  = p_2 - i\overline{\Pi}_1 \partial_2 \log a C^1
+ \frac{i}{2} \partial_2 \log a,
$$
the state \r{23} obeys the following Cauchy problem
$$
\frac{\partial}{\partial t}\Upsilon^t   =   [- \overline{\Pi}_1\Pi^1   -
\overline{\Pi}_2 \Pi^2  +  a  \mu_1  \partial_1  +  \mu_2 \partial_2 -
\frac{\mu_2}{2} \partial_2 \log a - \mu_1 \overline{\Pi}_1  \partial_2
\log a     \frac{\partial}{\partial    \overline{\Pi}_2}    +    \mu_2
\overline{\Pi}_1 \partial_2    \log     a     \frac{\partial}{\partial
\overline{\Pi}_1}] \Upsilon^t,
\l{24}
$$
$$
\Upsilon^0 = \Phi(q)
$$
Since eq.\r{24} is a first-order partial differential equation, it can
be solved by the characteristic method.  The solution is looked for in
the following form
$$
\Upsilon^t(Q^t,\tilde{\Pi}^t,\Pi) =         \exp[\int_0^t        d\tau
[- \tilde{\Pi}_1^{\tau} \Pi^1    -    \tilde{\Pi}_2^{\tau}    \Pi^2    -
\frac{\mu_2}{2} \partial_2            \log a(Q^{\tau})]               ]
\Upsilon^0(Q^0,\tilde{\Pi}^0,\Pi),
$$
where the  functions  $Q^t$,  $\tilde{\Pi}^t$  satisfy  the  following
ordinary differential equations,
$$
\dot{Q}_1^t = - a(Q_2,Q_3)\mu_1;  \qquad \dot{Q}_2^t = -\mu_2,  \qquad
\dot{Q}_3^t = 0,
$$
$$
\frac{d}{dt}\tilde{\Pi}_2^t =   \mu_1  \tilde{\Pi}_1  \partial_2  \log
a(Q_2,Q_3),
\qquad
\frac{d}{dt}\tilde{\Pi}_1^t = -  \mu_2  \tilde{\Pi}_1  \partial_2  \log
a(Q_2,Q_3),
$$
so that the classical characteristic trajectory is
$$
Q_3^t = Q_3^0, \qquad
Q_2^t = Q_2^0 - \mu_2 t, \qquad
Q_1^t = Q_1^0 - \int_0^t d\tau a(Q_2^0-\mu_2\tau, Q_3^0) \mu_1,
$$
$$
\tilde{\Pi}_1^t =   \frac{a(Q_2^0  -  \mu_2t,  Q_3^0)}{a(Q_2^0,Q_3^0)}
\tilde{\Pi}_1^0, \qquad
\tilde{\Pi}_2^t =  \tilde{\Pi}_2^0  +  \frac{1}{a(Q_2^0,Q_3^0)}  \mu_1
\tilde{\Pi}_1^0 \int_0^t d\tau \partial_2 a(Q_2^0-\mu_2\tau,Q_3^0).
$$
Combining all factors,  one finds the solution the Cauchy  problem
\r{24},
$$
\Upsilon^t(x,\overline{\Pi},\Pi) =
\sqrt{\frac{a(x_2,x_3)}{a(x_2+\mu_2t, x_3)}}   \exp[- \int_0^t
d\tau \frac{a(x_2+\mu_2t, x_3)}{a(x_2, x_3)} \overline{\Pi}_1 \Pi^1
- t\overline{\Pi}_2 \Pi^2]
$$
$$
\times
\exp[  \int_0^t d\tau \tau \mu_1
\frac{\partial_2 \log                    a(x_2+\mu_2\tau)}{a(x_2,x_3)}
\overline{\Pi}_1\Pi^2]
\Phi(x_1 + \int_0^t d\tau a(x_2+\mu_2\tau,x_3)\mu_1, x_2+\mu_2t, x_3)
\l{25}
$$
One can also check by the direct computations that  expression  \r{25}
really satisfies  the Cauchy problem \r{24}.  The inner product \r{15}
reads
$$
\int dx  \Phi^*(x)  \prod_{a=1}^M  d\mu_a   d\overline{\Pi}_a d\Pi^a
\Upsilon^t(x,\overline{\Pi},\Pi).
\l{26}
$$
Integration over ghost variables gives us the multiplier
$$
t\int_0^t d\tau a(x_2+ \mu_2\tau, x_3) \frac{1}{a(x_2,x_3)}.
$$
After rescaling of variables $\mu$
$$
\xi_1 = \int_0^t d\tau a(x_2 + \mu_2\tau,x_3) \mu_1, \qquad
\xi_2 = t\mu_2
$$
one finds that the integral \r{26} takes a simple form
$$
\int dx_1dx_2dx_3  d\xi_1  d\xi_2  \Phi^*(x_1,x_2,x_3) \frac{1}{\sqrt{
a(x_2+\xi_2,x_3) a(x_2,x_3) }} \Phi(x_1+\xi_1, x_2+ \xi_2,x_3)
$$
We see that the bilinear form $\eta$ can be defined as
$$
(\Phi,\eta \Phi)      =      \int      dx_3       |\int       dx_1dx_2
\frac{\Phi(x_1,x_2,x_3)}{\sqrt{a(x_2,x_3)}}|^2,
$$
so that  the correspondence between the Dirac wave function $\Psi$ and
the auxiliary state $\Phi$ is
$$
\Psi(x_1,x_2,x_3) = \frac{1}{\sqrt{a(x_2,x_3)}} \int dy_1 dy_2
\frac{\Phi(y_1,y_2,x_3)}{\sqrt{a(y_2,x_3)}}.
$$
It obeys the constraints
$$
a(x_2,x_3) p_1 \Psi \equiv \hat{\Lambda}_1^+ \Psi = 0, \qquad
\frac{1}{\sqrt{a(x_2,x_3)}} p_2    \sqrt{a(x_2,x_3)}    \Psi    \equiv
\hat{\Lambda}_2^+ \Psi = 0.
$$
while the gauge transformation of $\Phi$ is
$$
\Phi \to  \Phi + \sqrt{a(x_2,x_3)} p_2 \frac{1}{\sqrt{a(x_2,x_3)}} Y^2
+ a(x_2,x_3) p_1 Y^1
$$
for some functions $Y^1$ and $Y^2$.  We see  that  all  properties  of
$\eta$ (including positive definiteness)
are indeed satisfied in this example.

\section{Discussion}

Thus,  the refined algebraic quantization  formula  \r{4}  is
generalized to the case  of  structure  functions.  A  simple  exactly
solvable example  is  investigated;  an explicit formula for the inner
product is obtained.

A wide class of such examples of systems with structure functions  can
be constructed as follows. Consider the Lie-algebra constrained system:
$\hat{\Lambda}_a = L_a - \frac{i}{2} f^c_{ac}$, $U^a_{bc} = f^a_{bc} =
const$ such that $L_a$ are linear in momenta,
$L_a = \alpha_{aj}(x) p_j + \beta_a(x)$.
The B-charge is
$$
\Omega_0 =  C^a  L_a  - \frac{i}{2} f^a_{bc} \overline{\Pi}_a C^bC^c -
\frac{i}{2} f^a_{ba} C^b - i\pi_a\Pi^a.
$$
Consider the unitary transformation being an exponent of the  operator
quadratic with respect to ghost variables,
$$
U = \exp[\overline{\Pi}_a A^a_b(x) C^b - \frac{1}{2} A^a_a(x)]
$$
It generates   a   linear  canonical  transformation  of  ghosts.  The
transformed B-charge $U^{-1}\Omega_0 U = \Omega$ is Hermitian
and nilpotent. It contains terms $\Omega^1$ and $\Omega^2$ only
and corresponds to the new system with classical constraints
$$
\Lambda_{a'} = L_a (\exp A)^a_{a'}.
\l{27}
$$
with quantum corrections.
Generally, they form an algebra with nontrivial  structure  functions.
Since $\Omega^n=0$,  $n\ge3$,  while  the  constraints  are  linear in
momenta, the Cauchy problem analogous to \r{24} still  corresponds  to
the first-order  partial  differential  equation  and  can  be  solved
exactly, so that it is also possible  to  perform  an  integration  in
eq.\r{16} explicitly.

We see  that  the  system  with classical constraints \r{27} which was
mentioned in \c{Marolf3} can be exactly investigated by  the  approach
proposed in this paper.

The case   of  an  open  gauge  algebra  corresponding  to  nontrivial
coefficient functions $\Omega^n$,  $n\ge 3$ is much  more  complicated
for the  exact calculations.  However,  the integral formula \r{16} is
still valid,  so that one can use it for numerical calculations or for
application of  asymptotic  methods  such  as  perturbation  theory or
semiclassical approximation.

This work was supported by
the Russian Foundation for Basic Research, project 01-01-06251.

\end{document}